\documentclass[10pt, twocolumn]{article}
\usepackage{graphicx}
\usepackage{url}
\usepackage{upquote}

\begin{document}

\title{XTribe: a web-based social computation platform}

\author{ Saverio Caminiti$^1$, Claudio Cicali$^2$, Pietro Gravino$^3$,
			Vittorio Loreto$^{2,4}$,\\
    		Vito~D.~P.~Servedio$^4$, Alina S\^irbu$^2$ and 
    		Francesca Tria$^2$\\[3mm]
    		$^1$Institute for Complex Systems (ISC), CNR, Rome, Italy\\
			$^2$Institute for Scientific Interchange (ISI), Turin, Italy\\
			$^3$University of Bologna, Physics Department, Bologna, Italy\\
			$^4$Sapienza University of Rome, Physics Department, Italy
			}
			
\maketitle

\begin{abstract}
\textbf{In the last few years the Web has  progressively acquired the
status of an infrastructure for social computation that allows
researchers to coordinate the cognitive abilities of human agents in
on-line communities so to steer the collective user activity towards
predefined goals. This general trend is also triggering the adoption
of web-games as a very interesting laboratory to run experiments in
the social sciences and whenever the contribution of human beings is
crucially required for research purposes.
Nowadays, while the number of on-line users has been steadily growing,
there is still a need of systematization in the approach to the web as
a laboratory.
In this paper we present Experimental Tribe (XTribe in short), a novel
general purpose web-based platform for web-gaming and social
computation. Ready to use and already operational, XTribe aims at
drastically reducing the effort required to develop and run web
experiments. XTribe has been designed to speed up the implementation
of those general aspects of web experiments that are independent of
the specific experiment content.
For example, XTribe takes care of user management by handling
their registration and profiles and in case of multi-player games,
it provides the necessary user grouping functionalities.
XTribe also provides communication facilities to easily achieve both
bidirectional and asynchronous communication. 
From a practical point
of view, researchers are left with the only task of designing and
implementing the game interface and logic of their experiment, on
which they maintain full control.
Moreover, XTribe acts as a repository of different scientific experiments,
thus realizing a sort of showcase that stimulates users' curiosity,
enhances their participation, and helps researchers in recruiting
volunteers.%
\footnote{Please cite this paper as:\\
		S.~Caminiti, C.~Cicali, P.~Gravino, V.~Loreto, 
		V.D.P.~Servedio, A.~S\^irbu, F.~Tria,
\emph{XTribe: A Web-Based Social Computation Platform},
\textbf{IEEE Xplore}, Cloud and Green Computing (CGC), 
2013 Third International Conference on, 397-403 (2013). 
DOI:~10.1109/CGC.2013.69 }
}
\end{abstract}

\section{Introduction}
\noindent
Technology plays a fundamental role in connecting people and
circulating information, and affects more and more the way humans
interact with each other. The number of users surfing the Web exceeded
two billions in 2012 and an unprecedented huge amount of information
is everyday exchanged by people through posts and comments on-line,
tweets or emails, or phone calls as a natural aptitude of humans to
share news, thoughts, feelings or experiences. The Web is thus
entangling in an unpredictable way cognitive, social and technological
elements, giving rise in this way to the largest interconnected
techno-social system ever. Social networking tools allow effective
data and opinion collection and real-time information sharing
processes. The possibility to access the digital fingerprints of
individuals is opening tremendous avenues for an unprecedented
monitoring at a ``microscopic level'' of collective phenomena involving
human beings. We are thus moving very fast towards a sort of
tomography of our societies, with a key contribution of people acting
as data gathering ``sensors'' and with a level of fine-graining that
only two-three years ago would have been considered science fiction.
All this has deep implications for the understanding of the dynamics
and evolution of our complex societies as well as for our ability to
start making predictions and face the societal challenges of our era.
Social Science disciplines, traditionally depending on the recruitment
of test subjects to perform experiments, are for the first time
experiencing the possibility to gather significant data in a very
effective and capillary way, opening in this way the season of a
computational social science~\cite{lazer_2009}.

In this context, the use of the Web for research purposes is changing
the way research activities are conducted and how data are generated
and gathered in many scientific fields. Despite the prediction, cast
in 2009, that the new social platforms appearing on the Web might have
become a very interesting laboratory for social sciences in
general~\cite{lazer_2009}, Internet based research still lies in its
infancy and methodological and procedural obstacles have to be faced
in order to make it a reliable tool of investigation. Two paradigmatic
examples are \emph{Planet
Hunters\footnote{\url{http://www.planethunters.org}}}~\cite{PlanetHunter2011},
a game in which participants can help in identifying new extra-solar
planets using NASA data of star brightness and \emph{Galaxy
Zoo\footnote{\url{http://www.galaxyzoo.org}}}~\cite{galaxyzoo_2010},
in which players are asked to classify astronomic objects of galactic
type, by browsing a catalogue of telescopic images. The above mentioned
projects have in common the involvement of individual volunteers or
networks of volunteers, many of whom may have non specific scientific
training, to perform or manage research related tasks in scientific
projects. In this sense these are two examples of \emph{citizen
science}~\cite{ARNSTEIN69,Goodchild07a,COMMONSENSE09}, i.e., a
long-standing series of programs traditionally employing volunteer
monitoring for natural resource management.\\

\noindent Citizen science projects are becoming increasingly focused
on scientific
research~\cite{nosek_2002,salganik_watts_2009,foldit_2010} and amazing
results have already been obtained. For example, the 3D structure of
viral enzymes that challenged scientists for years has been discovered
thanks to the efforts of Foldit\footnote{\url{http://fold.it}}
players~\cite{foldit_2011}, new candidate planets identified by Planet
Hunters' participants managed to survive data verification
tests~\cite{PlanetHunter2011}, and brand new astronomical objects were
discovered
by Galaxy Zoo's users~\cite{galaxyzoo_2010}. 
These examples show how social computation
processes hold tremendous potential to solve a variety of problems in
novel and interesting ways, and how amateur players are able to solve
research problems, even faster than their professional researchers
counterparts. Human ability to easily solve tasks that are difficult
to solve by machines has been largely exploited for instance in
labelling images, through the collaborative \emph{ESP
Game}~\cite{vonahn_2004}, or in language automatic translators,
through the interactive learning platform
\emph{Duolingo}\footnote{\url{http://duolingo.com}}. In these last
two examples, the idea of linking playful activities with learning
processes has led to the paradigm of Games With a Purpose
(GWAP)~\cite{vonahn_2006}, i.e.\ a way of engaging people in games
that can extract valuable information or work as a side effect of the
game or the learning dynamics. The playful rearranging of experiments,
together with their appealing graphic interfaces, has also proved to
be a fundamental ingredient for web-based experiments design, boosting
user participation and data reliability.

\noindent This idea of \emph{crowdsourcing}, term coined in 2006
\cite{howe_2006}, is also at the heart of on-line labour markets such
as Amazon Mechanical Turk (AMT), where a job is distributed by
employers in small sub-tasks that on-line workers can perform in
return of proportionally small monetary payoffs. Interestingly,
despite its mercenary aspect, AMT has proven to be useful for
scientific purposes~\cite{chilton09,mason_watts_2009,Paolacci2010}, by
leveraging on its ease in recruiting a potentially large number of
experimental subjects. This early experience with crowdsourced
experiments has led to the recognition that Web experiments, despite
the unavoidable partial control on the way participants are recruited
and on the context in which tasks are executed, can be successfully
used to study human collective behaviour and cognition and can provide
elements of validation of experimental practices in the
Web~\cite{suri10}.

The tenets of social computation are being increasingly exploited, but
its use in the scientific community still lacks systematization. The
realization of a single project often requires substantial effort and
web-based experiments are still far from being standard research
tools. The lack of tools that can greatly simplify and standardize the
design of Web games and experiments is a major bottleneck in the
exploitation of such new research opportunities. For example, despite
its versatility, AMT has not been conceived as an experimental
platform, lacking dedicated infrastructures for the design of
experiments, while offering some visual tools to develop simple
interfaces. Experimentalists are left with the task of designing their
own software solutions to manage interactions among participants and
to build effective interfaces. Moreover, individual solutions to such
problems often remain isolated with little or no cumulative growth of
tools and solutions. Hence the need of a versatile platform to
implement web-based experiments with a very small coding effort. This
is the aim of XTribe\footnote{Already active and available at \url{http://www.xtribe.eu}},
 a general
purpose platform to carry on experiments in the form of web-games. The
word ``game'' is here intended as a real time interaction protocol
among few players implementing a specific task, as well as a synonym
of experiment on interactive behaviour.
By providing the scientific community with a general purpose platform
for social-computation and web-gaming, XTribe gathers otherwise
separate efforts to use Web resources for scientific purposes and
provides the community with a tool to design experiments on the Web,
from simple polls to more complex multiplayer games,
bypassing much of the ``hard work'', e.g.\ hosting, user registry
handling and user pairing/grouping, communication protocols,
exceptions handling, etc.

The aim of this paper is to describe the XTribe platform and
to provide the essential ingredients that would allow 
researchers to create, submit and mantain their own experiments with ease.

\section{Anatomy of a multiplayer online experiment}
The GWAP applications cited above show a vast variety of features and a very heterogeneous set of targets. But even these motley experiences have elements in common, beside the general idea of exploiting the force of the crowd. 
In order to introduce the necessary steps to build a GWAP,
in this section we shall analyze the structural and technical components of a generic GWAP, from an abstract point of view without going into detailed technicalities.
As a guide we shall consider here the structure of the ESP Game, the most famous GWAP. 
In this game, two players are asked to tag the same image, 
trying to match their tags.
They will input as many tags as they want until one tag is in common to both; then they move to the next image. 
Within a time limit of 2.5 minutes, the players have to agree on as many images as possible, 
to increase their score. 
The goal of the game from the experimenter perpective is to obtain realistic valuable tags for online images, to be used by search engines. 
We shall consider this game as a prototype that will make the analysis of the typical game components more clear.

 At one extreme of our abstract structure lies the \emph{developer}, i.e.\ a researcher willing to create a web experiment.
At the other end lies the \emph{community}, i.e.\ the ensemble of  users who will play the game.
Depending on the experiment, this can be a wide community or a subset filtered by age, gender, language, interests or even geographical location. In the case of ESP, those are the players who tag images. 
Developer and users are just the two ends of a complex structure and in the following
subsections we shall describe what lies in the middle and permits the execution of the game.

\subsection{The interface: interacting with the user\label{sec:IIA}}
\noindent
In the GWAP experiments there is a flow of information that, in most cases, 
starts from the user, e.g.\ in response to a given question 
(``how will the other player tag this picture?'', in the ESP Game case). 
Therefore, the application will need a user interface allowing players to insert their answers. 
The interface should be designed by researchers
with the goal of optimizing users' experience, 
ensuring an easy and enjoyable interaction. 
The user has to invest her time in paying attention to the application and  the entertainment itself offered by the interface can be a reward for the user interaction.
Moreover, a successful interface design will not only persuade the user to spend her time on the application but will also stimulate her to involve other people. 
A well designed interface should also help her in voluntary recruiting acquaintances, e.g.\ by leveraging on social networks features, such as tweets about the results, Facebook sharing of the results, etc. 
Even if the fanciness of the interface is crucial, the designer has always to keep in mind the biases introduced by the interface. Each kind of interaction introduces biases, even the simple fact that users are interacting through a computer. As we said, the reliability of the information gathered is a fundamental point. Thus, the impact of each bias introduced has to be carefully considered in order to find a good compromise between the reliability of results and the user experience.

\subsection{The server side logic and storage\label{sec:IIB}}
\noindent
Once the information has been gathered by the interface, in order to give feedback to users or results to the developer, it is very likely that some elaborations will be needed. So the application will need a logic elaboration part. 
In the ESP case, the logic component receives the tags from each of the two players, compares them and when a match is found, feeds the interface with a new image to be labeled. When the time is over, this component computes a score and sends it to both players. While the interface runs on the browser, i.e.\ on the user computer, the information processing should happen server side, in order to guarantee reliability and security (reducing the risk of failures, cheating and hacking). Moreover, the game logic may require complex computation involving data that the researcher cannot or does not want to make available to the user browser. Beside this, there is also a matter of control: the logic part has to be directly managed by the developer, the other end of our scheme. 
Hence, it should run on a machine under the developer control where all data generated by the experiment can be properly stored for further research and analysis. The logic part will also provide content for the application (e.g., pictures in the ESP Game). In other words, the logic part will take care of filling the interface with input and feedback, as well as of gathering results.

\subsection{The rest: technical but necessary issues\label{sec:IIC}}

The interface and the logic are the nearest neighbour of the user and of the developer, respectively. These two parts are the core of the application, the ``unique'' parts designed by researchers precisely on the project target. But the application itself it is still far from being complete. There are at least three missing fundamental parts: 
\begin{enumerate}
	\item a communication protocol between the two parts;
	\item a user handling system;
	\item an instance processing mechanism.
\end{enumerate}

\subsubsection{Communication}
The communication between the interface and the logic is potentially very difficult to implement. If we consider the simple case of a client initiating the communication by sending a message to a server, the solution is quite easy to carry out (e.g., with a HTTP request). But in case of more complex communication structures, such as bidirectional asynchronous client-server communication or, in multiplayer games, client-client communication, the implementation can be quite a difficult task requiring more sophisticated technologies (e.g., web-sockets).

\subsubsection{User handling} 
When dealing with users, a certain set of functionalities is likely to be useful such as user registration handling and profile management. At a basic level, it is a matter of security and reliability, because registration can provide a first filter against bots. Beside this, many experiments require a certain level of profiling of the users, to differentiate or group them depending on the gender, age, language, etc. On the other side, users may enjoy to see the result of their efforts, in the form of scores, ranks, etc. So they would prefer their ``player'' identity to be recorded by the game. Obviously, linked to this, there are also privacy issues: the developer has to guarantee to the user that his personal data will not be disclosed.

\subsubsection{Instances}
Once the interface has been prepared, the logic is running, they are communicating and the user is registered (if required), an instance of the game still has to be created, in order to allow the user to join the experiment. By instance we mean the single execution of the experiment task involving one or more users.
This management is relatively easy for single player games, but it becomes non-trivial in case of multiplayer games. A ``waiting room'' has to be implemented, in order to make the users wait for others to join.

These three parts have two things in common.
They are needed (if not all necessary they are at least all very useful) in almost every kind of web-application and  are not particularly influenced by the specific experiment or game. 
Hence, since this three parts are almost unrelated to the experiment, they are the most technical and dull to implement.
That is why a framework or, even better, a platform that can take care of these functionalities automatically would make it easier to create web experiments. This is where the XTribe platform comes in, to provide the technical ``middleware'' (i.e.\ Sec.~\ref{sec:IIC}) and allow the author of the game to focus on the game-specific interface and logic (i.e.\ Sec.~\ref{sec:IIA} and \ref{sec:IIB}). But the benefits of the XTribe platform are not limited to these.

\section{XTribe platform in detail}
\noindent
The XTribe platform has been designed with a modular structure so that most of the complexity associated to running an experiment is hidden into a Main Server (called Experimental Tribe Server or \emph{ET Server} for short). In this way most of the coding difficulties related to the realization of a dynamic web application are already taken care by the ET Server and the realization of an experiment should be as easy as constructing a webpage with the main utilities for it. There are different kinds of
users of the platform: the system administrator who runs the whole ET Server
and provides all the necessary API's for it; the experimentalists who
run individual experiments; and the players who participate
in one or more individual games.

On the XTribe platform each user/player interacts with one or more of the
available experiments/games. Each game is conceived by the game
developers/researchers who monitor the evolution through their local machines. Games have two components: the user interface (UI) and the logic - game manager (GM). The interface is what is visible to players, and will interact with them. The GM is represented by those functional parts that process the action of the players in order to implement coordination and specific game logics. 
\emph{These two components (the UI and GM) have to be developed by the researchers}, 
since they are highly dependent on the game itself.
XTribe mediates the communication between the two and hosts the game interface. 
The GM part of the game is hosted by the researchers on their own server. 
In this way they can directly collect the data in real time and have full control 
over the experiment progression. 
It is important to remark that XTribe does not store the data coming from the hosted 
experiments. All scientific data collected during an experiment can 
be conveniently stored by the GM, so that only the researcher who developed and 
published the experiment benefits of the outcome of his/her work. Beside this,
gathering data directly grants the opportunity to analyse them as
soon as they enter the system in real time.

\begin{figure}[!t]
\centering
\includegraphics[width=.48\textwidth]{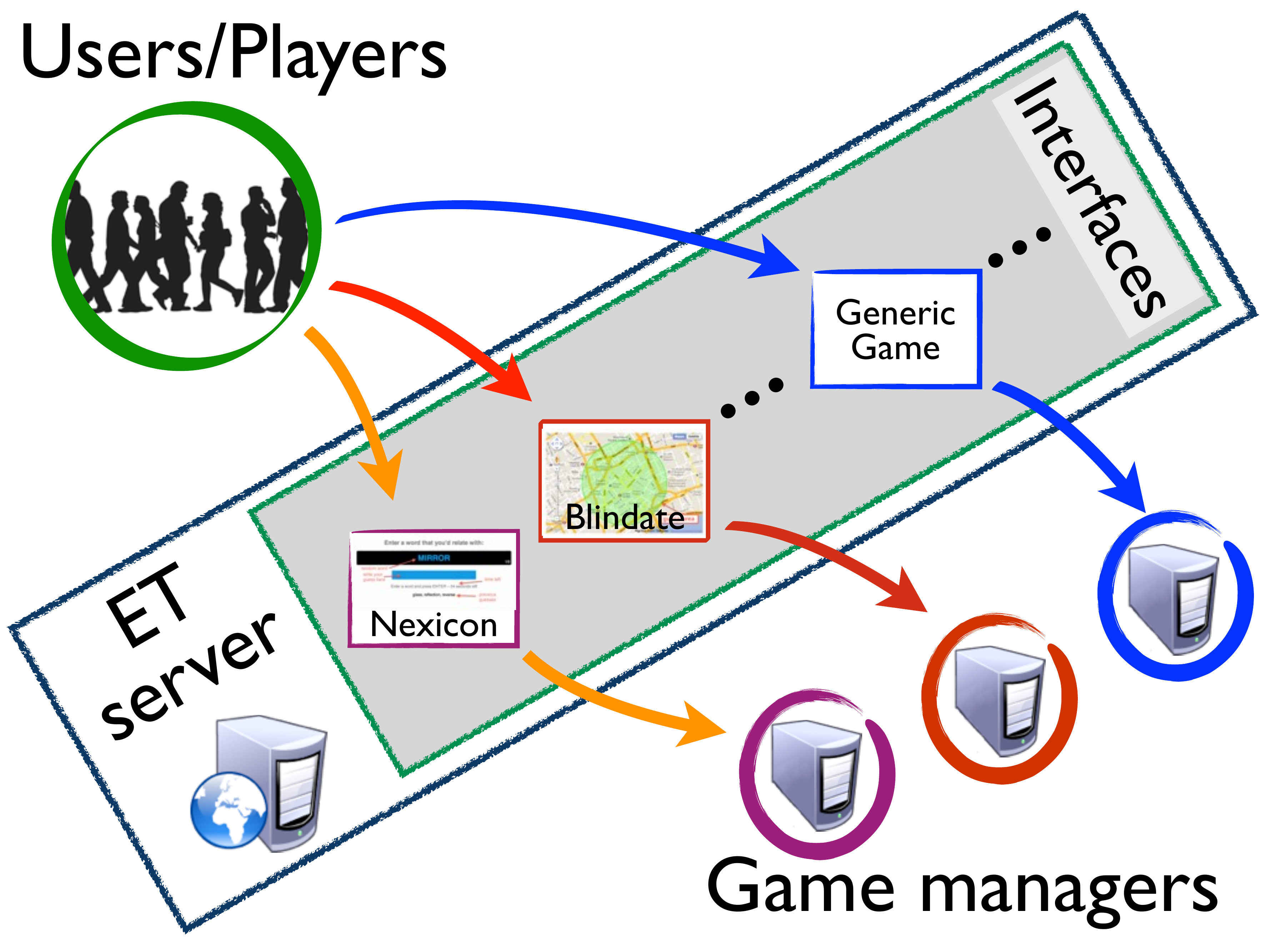}
\caption{A graphical representation of the system and its
  interactions. }
\label{fig:xtribe}
\end{figure}

The XTribe platform also offers a page for 
the description of the game rules, compiled by the researcher, from which 
players can access and play the game.
Additionally, it handles player/user management (registration,
authentication and profiling) and manages the actual instances of each experiment
(creation, user grouping, error handling, feedback to users and
managers, etc.). 
A graphical representation of the platform is depicted
in Figure~\ref{fig:xtribe}.

\subsection{User management and community}
\noindent
Since experiments are created for research purposes, the researchers
are interested in many types of statistics related to players. Beside
this, they may also be interested in filtering players for specific
purposes, e.g.\ according to their age, gender, language, geographical location,
etc. To this aim, XTribe handles a user registry in which
players will be allowed to register, if required, and play while the
system maintains all the information about them, such as scores,
ranks, game settings, leaderboards, etc.\ together with profile information. If needed, this
information can be sent to the GM, i.e.\ to the experimentalist.
Furthermore, based on this information, when properly configured, the
system will grant the access to the game only to certain profiles.
Being in charge of the handling of the user registry, the system would
also spare the researcher from dealing with privacy and security
issues since all data will be properly anonymized and, possibly,
encrypted. However, by default, it is still possible for unregistered users to
access the games. Filters are applied only if set by the researcher.

\subsection{Communication made easy}
\noindent
The communication between the UI and the GM is mediated by the ET Server through a message based protocol. The general functionality of a game can be summarised with the following flow:
\begin{itemize} 
\item Once the players have accessed the game, the system will create an
instance of the game. There may be given rules for the game to start.
A basic rule is the number of players. There may also be different constraints, e.g., pair players with similar scores or
players playing from different geographical locations. As soon as
there is a sufficient number of players satisfying the grouping
constraints, an instance of the game starts. 
\item The interface will transmit the actions
of the players to the GM, but all messages will pass through the
system, which will group them by match instance number after having
anonymized them. 
\item The GM will then receive the data, will elaborate
them and will send the results of the elaboration back to the system,
which in turn will transmit them to the UI of the various players. Obviously, the GM will
also save the data of interest locally (as it runs on the
researcher's machine).
\end{itemize}
It is important to remark that the GM can send messages to the UI either as a response of a message coming from a player (responding to that player, to the others or broadcasting to all of them) or by initiating the connection autonomously (e.g., after a given time).
The platform will also handle errors and
exceptions. For instance, if one of the
players disconnects unexpectedly, the system will detect and notify it
to the remaining players and will send a
message to the GM. Since there is no direct communication
between GM and interface, the GM will experience no trouble at
all.

\begin{figure}[!t]
\centering
\includegraphics[width=.48\textwidth]{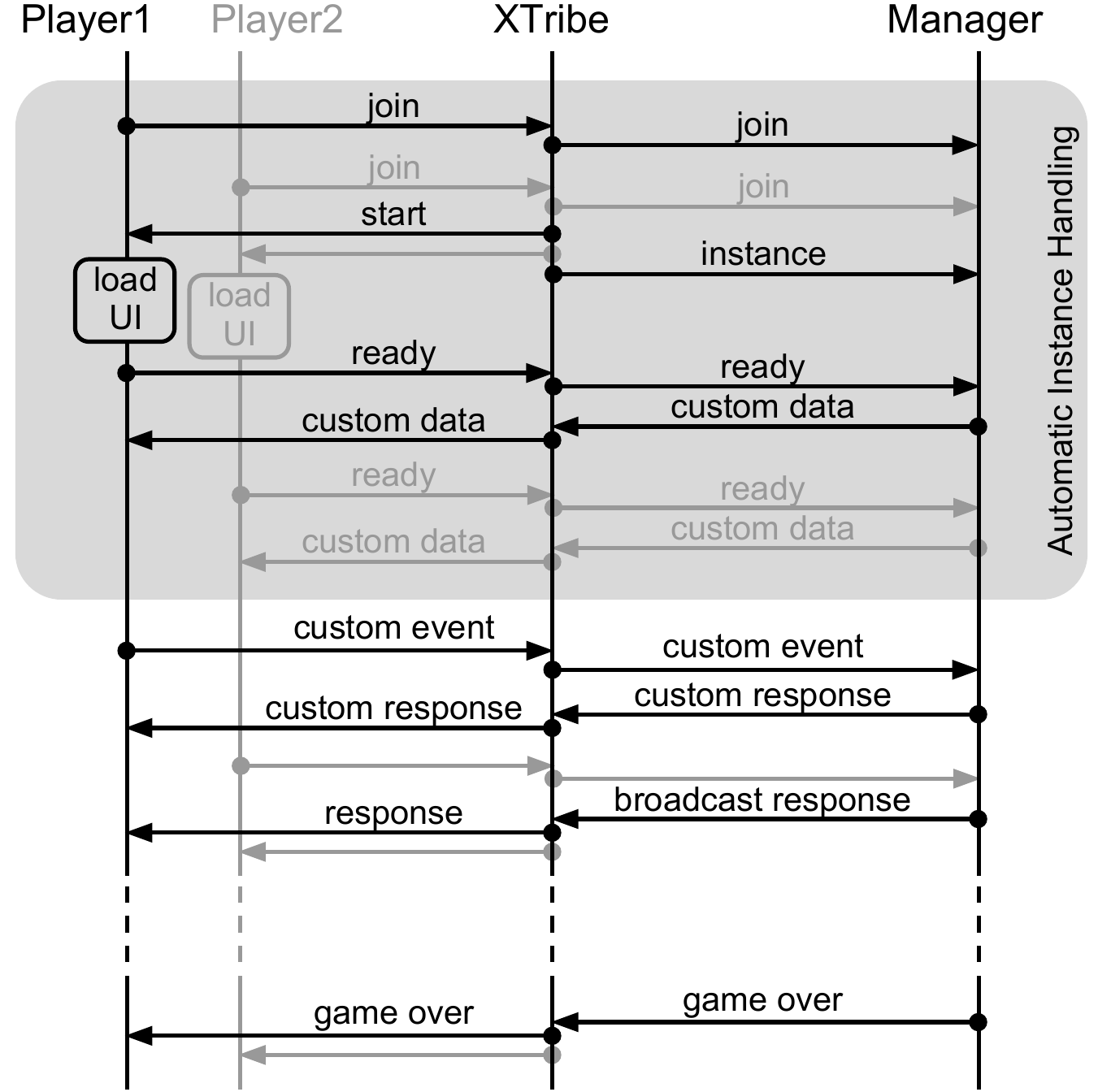}
\caption{The communication flow of a two-player game on XTribe. }
\label{fig:xtribecomm}
\end{figure}

In Figure~\ref{fig:xtribecomm} we depicted the communication flow of a two player game: a first player joins the experiment and waits for the second one to come. When both players are there an instance is created and the player's browsers are instructed to load the game UI. When loading is completed the UI notifies XTribe which in turn notifies the GM. Up to this point everything is automatic.
The GM will probably send custom data back to the players to let the game start. 
During the game custom data are exchanged between UI and GM, until the game is over and the instance is closed.

All these features, especially the user registry and the instance
handling, usually require a lot of coding, quantified in time and
money, to be realized. Within XTribe, they can be realized with a
straightforward procedure. After the configuration, the system will
automatically take care of all. What researchers have to do is to write
the code of the UI and of the GM only.

The UI has to be structured as a web page with plenty of
freedom in using HTML, CSS, Flash, etc., while the interaction between
the interface and the system has to be achieved by means of the ET Server
API, which are internally developed as Javascript functions. With this
simple set of functions the interface will interact with the platform
and, through it, with the GM. Basically, the GM has to work as a simple
HTTP server hosted on the researcher's machine. The communication with
the system takes place through the HTTP protocol and all messages are
coded in JSON format. The GM receives the message as a POST string variable and sends back one or more messages with a JSON string in the response body.
Besides a restricted set of system messages, the researcher is given full freedom to decide custom messages for the internal game protocol.

\subsection{Social network integration}
\noindent
Since the strength of online games comes from large participation, the XTribe platform has been integrated with the most powerful online social network application, \emph{Facebook}. Through Facebook the recruiting of new users is easier, since the new platform can spread through the network faster. The integration consists of the possibility to view the XTribe interface within the Facebook website and play games as Facebook games. Additionally, it provides seamless user registration, integrating the Facebook user information with the XTribe user registry. Hence, players have a better user experience connecting to XTribe without having to insert their information again, while researchers can collect more demographic information about the players of their games. Regular posts on user activity on the platform are published on user walls, and in this way additional players can be attracted to the system. Researchers wishing to build new games take advantage of this integration without any additional effort from their side. 

XTribe can be used in conjunction with the Amazon Mechanical Turk (AMT) platform in order to exploit its ability to 
recruit users with a modest monetary investment. AMT can be used to enhance participation and possibly in the initial phase of an experiment, to provide the necessary pool of data to begin with. The integration has been implemented by simply releasing an AMT payment code at the end of every single match or experiment. 

\section{XTribe for developers}
\noindent
The XTribe platform already features several experiments mainly about language, map perception, and opinion dynamics.  All scientists interested in developing experiments to be hosted on the platform can take advantage of the documentation and tutorials available on \url{http://doc.xtribe.eu}. Moreover they can be granted access to a sandbox version of the platform available at \url{http://lab.xtribe.eu}, where experiments can be tested during their development phase. All important XTribe related URLs are summarised in Table~\ref{tab:urls}.

\begin{table}
\begin{center}
\begin{tabular}{l|c}
\hline
	XTribe platform & \url{http://www.xtribe.eu}\\
	Documentation & \url{http://doc.xtribe.eu}\\
	Test platform & \url{http://lab.xtribe.eu}\\
\hline
\end{tabular}
\end{center}
\caption{Important XTribe related URLs.}
\label{tab:urls}
\end{table}

In the following we will briefly describe how implementing a multiplayer experiment on XTribe is a matter of hours, provided that the developers have basic knowledge of HTML, Javascript, and any server side language. We will use as a sample the Minority Game, described below.

\noindent{\bf Minority Game}. The game requires three players, who are presented with two choices (e.g., two numbers or two amounts of money). Each player has to choose one of the two options. If all of them agree on a single choice nobody wins and nothing happens. If only two players agree on their choice, they lose while the other player wins the amount they chose. A working version of the Minority Game can be played on \url{http://lab.xtribe.eu}.

In this very simple experiment a scientist may be interested in comparing user choices when numbers are shown with or without a currency sign or whether changing the ratio or the order of magnitude of the amounts results in behavioural shifts.  Implementing this as a web game from scratch would require a lot of effort in managing waiting rooms for users in order to group them 3 by 3. Moreover, the players make their choices asynchronously. This means that when the last player makes the choice the game terminates and the server has to contact the other two players with the result: handling server to client communication within standard HTML pages is not easy and requires efforts in the implementation and even in the server side configuration\footnote{Consider for example the fact that a user firewall may block communications on ports other than 80. To bypass this, all communication (including websockets) has to be routed through well known ports.}. On XTribe the experimenter is left with the only duty of 
implementing the HMTL interface and a server side script that, given a group of three players (identified by the platform with a unique instance number), chooses the two values to show, collects answers, and determines the winner (broadcasting the result to the players).

\noindent{\bf User Interface}. The implementation starts with creating an HTML page with two buttons, one for each choice. The developer then has to include, in the page head, the script {\tt Client.js} which makes ET Server API available to the experiment. The game UI will be hosted on the XTribe server where the API is.
In order to use the ET Server API it is enough to instantiate it:
\vspace{-1em}
\begin{verbatim}
c = new ETS.Client();
\end{verbatim}
\vspace{-1em}
and then register a user defined callback function to receive messages:
\vspace{-1em}
\begin{verbatim}
c.receive('manager', myFun); 
\end{verbatim}
\vspace{-1em}

The UI can receive messages both from the {\tt manager} or from the {\tt system} (especially useful to handle errors). Each message has a {\tt topic} (a string describing what the message is about) and a {\tt params} field which can contain arbitrary data about the message.
In our game the manager will send to the user two types of messages, one with the two choices at the beginning of the game, and one with the result at the end of the game. Consider these to be {\tt mgChoices} and {\tt mgResult} respectively.

\noindent A possible implementation of {\tt myFun} would be:
\vspace{-1em}
\begin{verbatim}
function myFun(msg) {
  switch(msg.topic) {
    case 'mgChoices':
      play(msg.params[0], msg.params[1]);
      break;
    case 'mgResult':
      answer(msg.params);
      break;
  }
});
\end{verbatim}
\vspace{-1em}
where {\tt play} and {\tt answer} are two user defined functions that change the HTML page according to the received values: the first one will fill the two buttons with the proper values chosen by the game manager, while the second one will show a message to the users depending on whether they win or lose.

The last thing that should be implemented on the UI is sending a message to the manager in response of a user interaction (i.e., button click). This can be easily achieved with ET Server API as follows:
\vspace{-1em}
\begin{verbatim}
c.send('manager', 'mgUChoice', v);
\end{verbatim}
\vspace{-1em}
where {\tt mgUChoice} is the arbitrary topic that describe this kind of message and {\tt v} is a variable that refers to what the user chose.

\noindent{\bf Game Manager}. The game manager runs on the experimenter server and can be implemented in any programming language (we use PHP in this example). It receives messages from XTribe as a POST variable exactly as a common script receives strings from an HTML form. The variable name is {\tt message} and it is a JSON encoded structure:
\vspace{-1em}
\begin{verbatim}
$msg = json_decode($_POST['message']);
\end{verbatim}
\vspace{-1em}

Looking at the {\tt sender} (either {\tt system} or {\tt client}) and {\tt topic} fields of the message, the manager will be able to take proper actions. Other relevant fields of the received message that have to be used are {\tt instanceId} and {\tt clientId}. These are two numbers generated by XTribe that univocally identify the instance this message refers to and the user who sent this message (if applicable).

In our game, as soon as three players joined the experiment, XTribe creates a new instance and notifies this to the manager with a message with {\tt sender = system} and {\tt topic = instance}. This is the perfect moment to generate the two values these three players will be playing with. These values can be stored, in association with the provided {\tt instanceId} in a database table or in some persistent data structure (easier for GM written with Java, NodeJS, Python).
The players will be loading the HTML interface in the mean time. As soon a each player is ready, this event is automatically notified to XTribe and in turn to the manager. The manager can then send a first message to the player, in our case a message with topic {\tt mgChoices} with the two values as {\tt params} (e.g., as an array). To send a message back to XTribe the manager simply writes it (as a JSON encoded string) in the body of the response: it is as easy as returning plain text:
\vspace{-1em}
\begin{verbatim}
$resMsg = array(
    'recipient' => 'client',
    'topic' => 'mgChoices',
    'clientId' => $msg.clientId,
    'instanceId' => $msg.instanceId,
    'params' => array(v1, v2)
  );
print(json_encode($resMsg));
\end{verbatim}
\vspace{-1em}

Each time the manager receives a message with topic {\tt mgUChoice} it stores the user choice updating the database or the persistent data structure. No response is required for the first two players, but when the third one answers the manager computes the winner and sends a broadcast {\tt mgResult} message back to all users plus an {\tt over} message to the system to inform it that this instance is over. Both messages can be sent together as an array.
\vspace{-1em}
\begin{verbatim}
$resMsg[0] = array(
    'recipient' => 'client',
    'topic' => 'mgChoices',
    'broadcast' => true,
    'instanceId' => $msg.instanceId,
    'params' => winner
  );
$resMsg[1] = array(
    'recipient' => 'system',
    'topic' => 'over',
    'instanceId' => $msg.instanceId,
  );
print(json_encode($resMsg));
\end{verbatim}
\vspace{-1em}
Optionally, with the {\tt over} message, the manager can provide a score for each player that will be used for the game leaderboard automatically managed by XTribe.

\noindent{\bf Deploy the game}. Once the UI and GM are ready the experimenters will simply create a new experiment on XTribe, providing basic information such as game name, description, icon, screenshots, number of players, etc.
Then they will simply upload all the UI files on XTribe and provide a URL to contact the GM running on their server.

\section{Conclusions}
\noindent
XTribe is a general purpose platform that handles all the aspects of the realization of web
experiments that do not concern directly the game itself. In this way, it
allows researchers to focus only on the core of the experiment,
leaving the rest to the system.

The platform is already running and has proven its usefulness with several games already implemented by different researchers. The already existing games refer to studies in language and opinion dynamics, where the human component plays a crucial role, and are designed as web based social experiments. They show the versatility of the platform and its ability to host experiments on a diverse range of topics, as words association games, citizen mapping, response of individuals to traffic information, expressing political opinions. These are prototype experiments 
where issues concerning different aspects related to results reliability and to the recruitment ability of the platform, as well as of single games, can be addressed.
Besides their immediate scientific interest,
they are meant to open the way to  the use of this online laboratory, also involving other potentially interested research groups. In the immediate future, the platform will also host an air pollution mapping game that will be part of an international competition related to environmental awareness.   

An important result of the project is to allow researchers working in
different fields, who lack computer science expertise, to create
web-based experiments and games. In order to further facilitate this, the
next step is to create a set of ``default'' GMs for games
corresponding to the most standard types of web experiment, such as
surveys or coordination games. For the time being, there is a default
GM available that broadcasts to all the players the messages received
from each one.

As already stressed,  the platform is expected to act  as a reference point for interested users,
giving a fundamental boost in facing a
 typical issue related to web experiments: the recruitment. It is often quite difficult to gather a critical mass of ``suitable'' players, and this can be an easier task for an organized and collective platform than for single games. A first step towards facilitating recruitment was Facebook integration. In time, this process will become easier for new games. Since they are hosted on the platform, and shown on its main page, other players already involved in other games would probably join, attracted by curiosity. We expect a community of players to gather around XTribe playing different games and also giving researchers feedback about their experiments. We also expect researchers to aggregate into communities, sharing advices and best experimental practices with each other.
In the near future, the platform will made available classic tools for cooperation such us forum, to discuss experimental procedures, and a repository for GM and UI, where willing researchers can make their own code free for download and reuse.

\section*{Acknowledgment}
\noindent
We acknowledge useful discussions with M.~Warglien and G.~Paolacci at the early stage of the platform.
We acknowledge the EveryAware European project nr.\ 265432 under FP7-ICT-2009-C for financial support.


\end{document}